\documentclass[useAMS,usenatbib]{mn2e}

\newcommand{\msun}{M$_{\sun}$}

\newcommand{\kmss}{km s$^{-1}~$}
\newcommand{\nbody}{{\it n}-body$~$}
\newcommand{\afreeze}{$a_{frz}$}

\newcommand{\mnras}{MNRAS}
\newcommand{\apj}{ApJ}
\newcommand{\apjl}{ApJ}
\newcommand{\aj}{AJ}
\newcommand{\aap}{A\&A}
\newcommand{\araa}{ARA\&A}

\newcommand{\apjs}{ApJS}

\usepackage{graphicx}
\usepackage{amssymb,amsmath}
\title[Permanent soft binaries in dispersing clusters]{The formation of permanent soft binaries in dispersing clusters}
\author[N. Moeckel and C.J. Clarke]{Nickolas Moeckel$^{1}$\thanks{E-mail:
moeckel@ast.cam.ac.uk} and Cathie J. Clarke$^{1}$\\
$^{1}$Institute of Astronomy, University of Cambridge, Madingley Road, Cambridge, CB3 0HA\\
}
\begin{document}

\date{Accepted XXX. Received YYY; in original form ZZZ}

\pagerange{\pageref{firstpage}--\pageref{lastpage}} \pubyear{2009}

\maketitle

\label{firstpage}

\begin{abstract}

Wide, fragile binary stellar systems are found in the galactic field, and have recently been noted in the outskirts of expanding star clusters in numerical simulations. 
Energetically soft, with semi-major axes exceeding the initial size of their birth cluster, it is puzzling how these binaries are created and preserved.  We provide an interpretation of the formation of these binaries that explains the total number formed and their distribution of energies.  A population of weakly bound binaries can always be found in the cluster, in accordance with statistical detailed balance, limited at the soft end only by the current size of the cluster and whatever observational criteria are imposed.  At any given time, the observed soft binary distribution is predominantly a snapshot of a transient population.  However, there is a constantly growing population of long-lived soft binaries that are removed from the detailed balance cycle due to the changing density and velocity dispersion of an expanding cluster.  The total number of wide binaries that form, and their energy distribution, are insensitive to the cluster population; the number is approximately one per cluster.  This suggests that  a population composed of many dissolved small-{\it N} clusters will more efficiently populate the field with wide binaries than that composed of dissolved large-{\it N} clusters. Locally such binaries are present at approximately the 2\% level; thus the production rate is consistent with the field being populated by clusters with a median of a few hundred stars rather than a few thousand.
\end{abstract}

\begin{keywords}
binaries:general--methods:{\it N}-body simulations--stars:formation--stellar dynamics
\end{keywords}

\section{Introduction}
A large fraction of stars, across all masses, are found in multiple systems \citep[e.g.][]{duquennoy91,fischer92,sana10}.  The most recent survey of our local volume within 25 pc finds almost half of nearby solar-type stars \citep{raghavan10} are multiple.  The formation of multiples is presumably tied to the star formation process, and reproducing the statistics of stellar multiplicity is a necessity for a complete star formation theory.  However binaries, especially energetically soft binaries with binding energies low compared to the mean kinetic energy of their environment, are subject to modification due to encounters with other stars \citep{heggie75,hills75}.  The picture is thus complicated somewhat by the dynamical processing that takes place in a clustered birth environment \citep[e.g.][]{kroupa01,parker09a}.  

It is particularly interesting to consider very wide binaries, with separations comparable to or larger than a typical star forming region (a few $10^4$ to $10^5$ AU).  If they are dynamically created, it is unclear how such very soft pairs could survive a clustered birth; in a long lived cluster the soft binaries will be destroyed by repeated encounters with their cluster siblings. 

If the binaries form from the fragmentation of a single collapsing core, the upper limit on the binary separation should be roughly the Jeans length, typically $\sim 10^4$ AU, and even this is unlikely since it implies the fragmentation of a core that is already rotating near its break up velocity before collapse.  At separations above a few $10^5$ AU, the Galactic environment begins to impose an upper limit on observed binaries via encounters with field stars \citep{retterer82,weinberg87}, GMCs \citep{mallada01}, and at larger separations the Galactic tidal field \citep{jiang10}.   \citet{kouwenhoven10} provide a nice summary of current observations and the apparent difficulty of producing binaries with very wide separations.  The bottom line is that there is a window in the range of about $10^4$ to $10^5$ AU where binaries are observed, and whose formation is difficult to explain.

The puzzling production of these wide binaries is the focus of this paper.  Recently, two numerical studies noted the formation of very wide systems in dispersing clusters.  \citet{moeckel10}, advancing the hydrodynamic cluster formation model of \citet{bate09} with an \nbody code after removing all the gas, found that some systems with semi-major axes $> 10^4$ AU formed in the expanding halo.  While the hard-soft borderline separation becomes larger as the cluster expands, this is not the solution;  the wide binaries that form in simulations and are seen in the field are soft even by the more generous standards of their dispersing natal clusters.

\citet{kouwenhoven10} specifically explored wide binary formation in expanding clusters, using \nbody models of clusters from both virial and super-virial initial conditions, and from fractal and spherical initial conditions.  They found in their final binary distribution two peaks in semi-major axis; one associated with the hard binaries that form at core collapse and drive the expansion of an isolated cluster, and another at much larger separations that they termed the dynamical peak.  They proposed that this peak forms as stars that are randomly associated in the phase space of a Maxwellian distribution find themselves bound as the cluster potential becomes less important in the halo of an expanding cluster.

In this paper we offer an alternative explanation for the creation of a population of soft yet permanent binaries during the expansion of a cluster.  The instantaneous soft binary population in a cluster reflects a statistical detailed balance, as pairs of stars are perturbed into and out of formally bound states.  Any single soft binary is typically short-lived, though the overall population is approximately static.  The expansion of the cluster and the consequent lowering of the stellar density leads to a constantly changing environment in which the binaries find themselves.  Binaries that are energetically soft by local standards at the time of their formation, with semi-major axes similar to the interstellar separation, can survive if the timescale on which the density drops is sufficiently short compared to the disruption timescale of the binary.  These binaries, effectively frozen out of the binary creation-destruction cycle, become a population of energetically soft, yet permanent binaries.

We begin in section \ref{simsection} by describing a suite of \nbody simulations performed to explore this idea, and by discussing the properties of the resultant binary population.  In section \ref{expsection} we develop our explanation of the freeze-out process, augmented by the appendices, and in section \ref{conclsection} we discuss the implications of this work for generating a field population of wide binaries before summarising our conclusions in section \ref{conclusions}.

\section{The Simulations Described}
\label{simsection}
\subsection{Numerical setup and method}
We simulated isolated star clusters using the GPU-enabled version of {\sc NBODY6} \citep{aarseth00,aarseth03}, largely following some of the setups used in \citet{kouwenhoven10}.  Our initial density structure was a Plummer sphere \citep{plummer11} in virial equilibrium, and we took our masses from a \citet{kroupa02} distribution in the range 0.08--8.0 \msun.  We ignored any tidal effects and stellar evolution, we have no primordial binaries, and we evolved the clusters using standard \nbody units \citep{heggie86} where the total mass, gravitational constant, and virial radius are related by $M = G = R_v=1$.  In these units the initial cluster crossing time is $2 \sqrt{2}$, and the cluster energy is -0.25. The absolute value of the specific binding energy of a binary, $\epsilon$, is unity at the borderline between a hard and a soft binary, with very soft binaries having $\epsilon \ll 1$.

We modeled clusters with {\it N=} 256 and {\it N=} 1024 (1k) for $5 \times 10^4$ time units, and 8196 (8k) stars, for $10^5$ time units.  For clarity of presentation, we focus mainly on the 1k runs.  To gain statistical leverage, we ran 48 runs at  256 and 1k and 12 runs at 8k; when discussing bulk cluster properties we are referring to the median values over all runs, and binary population distributions use the total binary population summed over all runs.  To identify wide binaries, we first find the nearest neighbour of each star.  If two stars are mutually nearest neighbours, we calculate their energy relative to their center of mass, and those pairs that have negative energy are deemed binaries.

While we present most of our results in \nbody units, the scaling of \nbody simulations to physical units is straightforward.  Physical lengths are obtained by simply multiplying all length scales by the initial virial radius of the cluster.  Our choice of a mass function give a mean stellar mass $m \approx 0.45$ \msun.  With mass measured in Solar masses and distances in parsecs, characteristic velocities in \kmss are obtained by $\sigma \approx 0.066 ( N m / R_v )^{1/2}$, and one unit of time is converted to Myr with the factor $15 (N m / R_v^3)^{-1/2}$.  Choosing either an initial velocity dispersion of a few \kmss or an initial virial radius of the order 0.1 pc are equivalent, and when we discuss results in physical units we choose $R_v = 0.1$ pc as an illustrative case.

\begin{figure}
 \includegraphics[width=80mm]{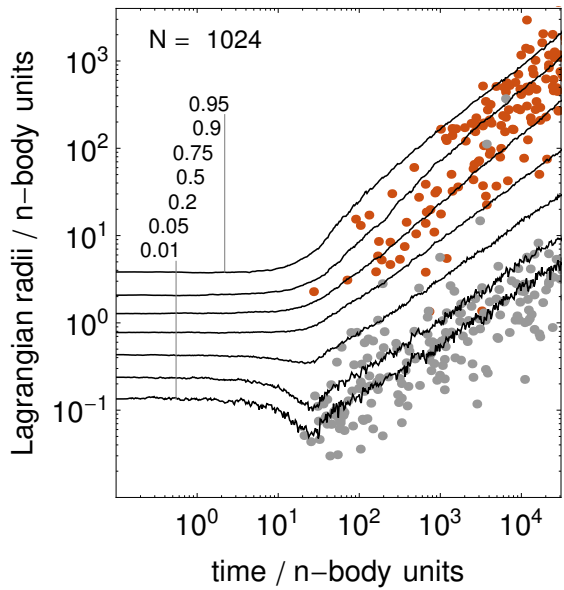}
 \caption{Lagrangian radii for the 1k clusters, shown as the median value of the 48 runs; solid lines show the radius enclosing the labeled mass fraction. The formation time and location of all binaries present at the end of the simulations are shown as dots.  Soft binaries, with $\epsilon<1$, are shown in orange, and hard binaries with $\epsilon>1$ are shown in grey.}
 \label{1024lagr}
\end{figure}

\subsection{Cluster expansion}
The clusters undergo relaxation driven global expansion as a result of stellar dynamical effects.  All the results we describe here may not apply to expansion from other processes, such as gas removal.  The overall structure of the clusters proceeds along expected lines, which we describe briefly.
In figure \ref{1024lagr} we show the evolution of the Lagrangian radii for the 1k runs.  Lagrangian radii enclose a fixed mass fraction relative to the cluster density center, which is determined via the prescription of \citet{casertano85}.  We exclude escaping\footnote{ For these purposes, any star outside the 0.9 Lagrangian radius with a velocity exceeding twice the escape velocity at its current radius is deemed to be an escaper.} stars from the calculation of the Lagrangian radii, although they are not removed from the simulation and can contribute the binary counts. The long-term evolution of these radii is well-established \citep[e.g.][]{baumgardt02}.  The early evolution is marked by contraction of the innermost Lagrangian radii, as two-body relaxation conducts kinetic energy outward from the core, which responds by contracting in a vain attempt to find equilibrium \citep{lynden-bell80}.  The evolutionary timescale is approximately the half-mass relaxation time,
\begin{equation}
t_{rh} = 0.138 \frac{N^{1/2}R_h^{3/2}}{(G m)^{1/2}{\rm ln} (\gamma N)},
\end{equation}
where $R_{h}$ is the half-mass radius, $N$ the number of stars, and $m$ the mean stellar mass.  We take as the argument in the Coulomb logarithm $\gamma = 0.02$, appropriate for our mass function \citep{giersz96}.  Core collapse of a Plummer sphere with equal masses takes place after about 15 $t_{rh}$, although the presence of a mass function accelerates the collapse as mass segregation drives the massive stars to the core.  Once there, the mass-segregation instability \citep{spitzer69} drives the core to collapse faster than in the equal-mass case, and core collapse occurs at less than a relaxation time \citep{gurkan04}.

\begin{figure*}
 \includegraphics[width=160mm]{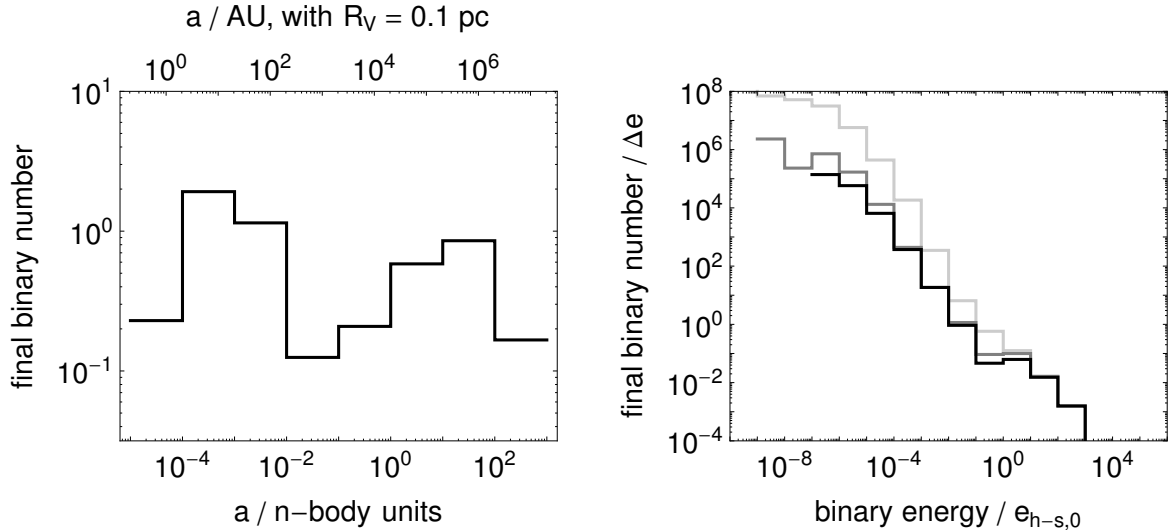}
 \caption{Semi-major axis and binding energy distribution of the binary population for the 1k runs.  The semi-major axis plots shows only the permanent binaries.  The energy plot shows the statistical population in light gray, the instantaneous population in medium gray, and the permanent population in black.  These plots are normalized by the total number of runs. For reference, the upper scale on the semi-major axis plot gives the values for a cluster with an initial virial radius of 0.1 pc; this can be linearly scaled to any preferred cluster size.}
 \label{1024semidist}
\end{figure*}

For our 1k systems, the initial relaxation time is $t_{rh,0} \approx 32$.  Core collapse occurs at about $t_{cc} \approx 27$, after which binaries in the core drive the expansion by heating their surroundings as they harden.  As two-body relaxation conducts this heat outwards, the cluster expands in response on the relaxation timescale.  This expansion is approximately self-similar, and the post core-collapse evolution can be reasonably described by 
\begin{equation}
r_{L} = r_{L,0} \left(1 + \chi \frac{t -t _{cc}}{t_{rh,0}} \right)^\delta,
\label{rlevolution}
\end{equation}
where $r_L$ is some Lagrangian radius.  For the half-mass radius itself, one can analytically show that  $\delta=2/3$, and while this value fits well interior to the half-mass radius, the outer radii expand with a slightly steeper power over the length of time we simulate.  The factor $\chi$ depends on the cluster mass function; \citet{gieles10b} give $\chi \approx 0.1(m_{max}/\bar{m})^{0.7}$, which we use here to find $\chi \approx 1.15$.


\subsection{Binary properties}
In figure \ref{1024lagr} we plot, in addition to the Lagrangian radii, the time of formation and radius in the cluster of all the binaries that are present at the end of the simulation.  While we identify the final binaries at the end of the simulation ($t = 5 \times 10^4$ for the 1k runs), we only plot those binaries that formed before $t = 2.5 \times 10^4$, guaranteeing that they are truly long-lived binaries and not a transient population. This population of binaries is referred to throughout the paper as `permanent'.  There are two clear populations; those that form in the center of the cluster and that are supplying the energy to drive the cluster expansion, and a class of wide binaries that form in the outskirts of the cluster.  These are the soft binaries that primarily concern us here.  

In figure \ref{1024semidist} we plot the distribution of semi-major axes for the permanent binaries, and the spectrum of binary energies for three sets of binaries.  The black energy histogram shows the permanent binaries; the medium gray histogram shows all binaries (i.e. bound nearest neighbours) at $t = 2.5 \times 10^4$ regardless of their eventual permanence or destruction (the `instantaneous' binary population); and the lightest gray shows the distribution of all bound pairs at $t = 2.5 \times 10^4$ regardless of their proximity (the `statistical' binary population).  The latter two populations will be discussed in the following section.  The final distributions are characterized, as \citet{kouwenhoven10} pointed out, by a peak of hard binaries at small semi-major axes, and a peak at large separations which is the focus of that work and this one.  The borderline between a hard and soft binary is $a_{h-s} \sim R_v N^{-1}$, and so evolves with time in the same sense as the cluster expansion.  When plotted as their energy spectrum they are revealed as a distribution that rises smoothly towards low binding energy.

The relation between these three populations can be seen in figure \ref{1024semidist}. The statistical population extends to the lowest energies, bounded only by the size of the cluster.  The instantaneous population is further restricted by the definition that the stars need to be nearest neighbours.  At late times the difference between the instantaneous and permanent populations is only clear at the softest energies. 

\section{Explaining the Soft Binary Population}
\label{expsection}
\subsection{Freezing-out of detailed balance}
In a star cluster with a Maxwellian velocity distribution, any two stars have a statistical chance of having instantaneously negative relative energy.  In an infinite, uniform distribution of stars, there is a population of soft binaries at all energies, maintained by detailed balance as stars are perturbed into and out of bound states.  While an individual binary is a transient object, the statistical distribution is steady; this has been known for a long time \citep{heggie75}.  The simple statistics that describe the very soft energies become complicated when one moves closer to the hard-soft border, but the complete steady-state distribution of binaries in a cluster with a uniform, static density was thoroughly described by \citet{goodman93a}.  In that idealized density distribution, a small fraction of the soft binaries are perturbed into hard binaries, and become permanent.  The statistically constant supply of transient soft binaries provides a source of hard binaries, though individual soft binaries tend to be short lived.

In the isolated, modest-{\it N} clusters we have simulated here, the expansion of the cluster provides another route toward permanence for a soft binary.  As the cluster expands and the density drops, a binary that is soft by local standards can find itself isolated from the perturbations that would destroy it in a static cluster.  Provided the density drops on a timescale short compared to a binary's desctruction timescale, it is possible for an individual soft binary to drop out of the creation-destruction cycle and become permanent.  Given the details of the cluster expansion we can calculate this freeze-out binary separation \afreeze, below which binaries are likely to survive.  We detail this calculation in Appendix \ref{FreezeOutAppendix}.  

Another necessary criterion for potential permanence is that a soft binary have a semi-major axis that is less than about the mean interstellar separation $a_{sep}$ at the location of its creation.  A system with a binary separation larger than the local packing of stars is subject to constant perturbation rather than the discrete perturbations a smaller binary experiences.  The conditions are ripe for soft, permanent binary formation at separations $a$ when \afreeze $ > a_{sep} \sim a$.  As the cluster's structure evolves, the permanent soft binaries that form will be characterized by those regions of the cluster that meet these conditions.

\begin{figure}
 \includegraphics[width=80mm]{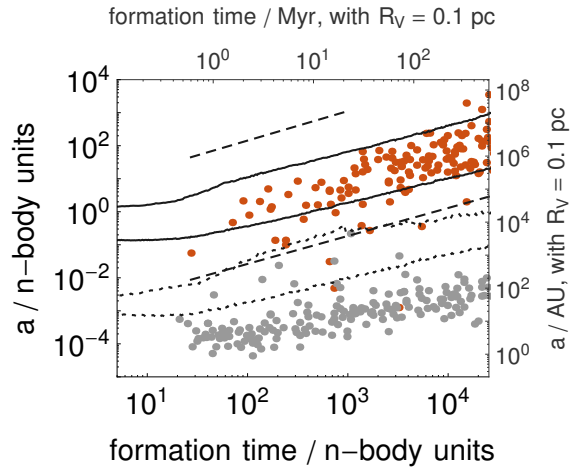}
 \caption{The formation time and semi-major axis of all permanent binaries from the 1k simulations.  In gray are energetically hard binaries and in orange are the soft binaries, matching figure \ref{1024lagr}.  The lines show three quantities at the 0.95 and 0.5 Lagrangian radii, with the 0.95 lines above the 0.5 lines.  Solid lines show the mean interstellar separation; dashed lines show the freeze-out radius from equation\ref{freezevalue}; and dotted lines show the hard-soft boundary for stars with the mean stellar mass.
 }
 \label{1024freeze}
\end{figure}

To show the relationship between the characteristics of the permanent binaries and the evolving cluster, in figure \ref{1024freeze} we plot the final semi-major axis of the permanent binaries against the time when the pair first became associated.  We also show \afreeze~for the 0.95 and 0.5 Lagrangian radii as dashed lines, over the range of time during which the density and velocity dispersions are reasonably approximated by a power law (figure \ref{1024densdisp}).  The 0.95 freeze-out radius, which is the upper dashed line, is well above any of the binaries that form as well as the interstellar separation\footnote{After the power-law behavior of the outer cluster is broken, the freeze-out radius becomes even less of a constraint due to the fact that the velocity dispersion falls off more slowly with time.}, and is about 3 orders of magnitude larger than the hard-soft semi-major axis at that radius, shown by the upper dotted line.

We show the freeze-out radius at the 0.5 Lagrangian radius, because very few permanent soft binaries form inside the half-mass radius (figure \ref{1024lagr}).  This is shown as the lower dashed line, only about 1 order of magnitude greater than the hard-soft boundary at that radius(the lower dotted line).  Throughout the evolution of the cluster, \afreeze~ at the half-mass radius is smaller than $a_{sep}$ by roughly an order of magnitude.  The conditions for binaries to drop out of the statistical balance are not met, and thus very few permanent soft binaries form interior to the the half-mass radius.  In the outer reaches of the cluster $a_{sep} <$\afreeze~ for the entirety of the cluster's evolution, and the semi-major axes of the permanent soft binaries formed tracks the evolution of $a_{sep}$.

\subsection{Estimating the number of systems that form}
In this picture, in regions where $a_{sep} <$\afreeze~ most nearest neighbours that are instantaneously bound should stay bound permanently.  In order to estimate the number of soft permanent binaries that form, we need to determine the fraction of  nearest neighbours that are bound.

We begin by estimating a characteristic interstellar separation at the virial radius $R_v$ as $a_{sep} \approx R_v N^{-1/3}$.  The velocity associated with this separation is $v_{sep} \approx (G m / a_{sep})^{1/2}$.  The boundary semi-major axis between a hard and soft binary is given by $a_{h-s} \approx R_v N^{-1}$, and is related to the velocity dispersion by $\sigma \approx (G m / a_{h-s})^{1/2}$. 

If the nearest neighbour of a star is at something close to the interstellar separation, what is the fraction of stars in the cluster that have a relative velocity low enough for them to be bound?  The relative velocities are a Maxwellian with dispersion $\sqrt{2} \sigma$, and we are interested in pairs with velocities below $\sqrt{2} v_{sep}$.  The fraction of bound neighbours is the integral over the Maxwellian velocity distribution up to that maximum velocity,  
\begin{equation*}
f_{bound} = \frac{1}{2 \sqrt{\pi}}\int_0^{v_{sep}/\sigma} e^{-x^2/4} x^2{\rm d}x,
\end{equation*}
where we are integrating over the ratio of the velocity to the dispersion.
The ratio $v_{sep} / \sigma \approx N^{-1/3}$, even for clusters as small as $N \sim 100$ less than about 0.2.  In this limit the exponential term is very close to unity, and the bound fraction is well approximated by
\begin{equation*}
f_{bound} \approx \left(\frac{v_{sep}}{\sigma}\right)^{3}.
\end{equation*}

As noted, that ratio of velocities is approximately $N^{-1/3}$, so the bound fraction should be $f_{bound} \approx N^{-1}$.   A point of concern is that we are calculating using a single value of these variables for the entire cluster, and taking those values at the virial radius rather than the outer regions of the cluster.  However, when the expansion of the cluster is driven by relaxation and nearly homologous, the local velocity dispersion tracks with the virial velocity, offset by a factor of order unity (see figure \ref{1024densdisp}).  Also note that the final result boils down to the ratio of $v_{sep}$ to $\sigma$, which can be recast as a ratio of $a_{h-s}$ to $a_{sep}$.  The ratio of these two quantities is roughly constant over the entirety of the simulations at all radii, which can be seen in figure \ref{1024freeze}.  The correction to the bound fraction calculation due to variations at different radii will be a factor of order unity, not affecting the scaling.

This argument has implications for the final binary energy spectrum and semi-major axis distribution.  Provided there is some region where $a_{sep} <$\afreeze, as the cluster expands and $a_{sep}$ grows larger the permanent binary population is built up from small separations up to large ones.  If we consider a range of semi-major axes around the critical separation at a given time, say a factor of 2, the number of bound pairs will be roughly $f_{bound} N$, or of order unity.  Thus as the region where soft permanent binaries are created sweeps through larger values, seen in figure \ref{1024freeze} as the region between the black lines, the number of permanent binaries formed in any logarithmic separation bin should be of order unity.  Note that this is independent of the population of the cluster.  As long as there is a region of the cluster producing soft permanent binaries, details of the cluster density structure and population do not matter.

A flat distribution in the logarithm of the semi-major axis means that the energy spectrum should follow ${\rm d}N_B/{\rm d}\epsilon \propto \epsilon^{-1}$.  These scalings are roughly born out by our simulations; the best fit to the permanent binary spectrum over the soft energies in figure \ref{1024semidist} is  ${\rm d}N_B/{\rm d}\epsilon \propto \epsilon^{-1.2}$, and the number of permanent soft binaries in logarithmic bins of semi-major axis are within a factor of a few over four orders of magnitude.  

The permanent soft population is a fossil record of the widest interstellar separation achieved by the regions of the cluster from which soft binaries are able to freeze out.  The final population is independent of the cluster parameters, such as the detailed density structure.  This is in marked contrast to the energy spectra of the statistical and instantaneous populations, which are functions of the finite size and geometry of the cluster.  The steady-state statistical spectrum derived in a uniform, infinite cluster is ${\rm d}N_B/{\rm d}\epsilon \propto \epsilon^{-5/2}$ at soft energies \citep{heggie75,goodman93a}.  When the cluster is finite, this spectrum is truncated at the soft end by the size scale of the cluster, which defines the largest possible separation of two stars, and thus the lowest energy binary attainable.  

When the density field is non-uniform, the slope of the statistical energy spectrum changes.  For instance, a star sitting at the center of a cluster with a density profile decreasing outwards will see fewer potential binary partners at large separations compared to a uniform distribution (see appendix \ref{DensityAppendix}), which leads to a flattening of the energy spectrum relative to the uniform case.  Both of these effects can be seen in figure \ref{1024semidist}, where the statistical population has a slope at soft energies more like ${\rm d}N_B/{\rm d}\epsilon \propto \epsilon^{-1.5}$, flattening further at the softest energies where it is eventually truncated.  The instantaneous spectrum, which is some subset of the statistical population, will likewise reflect the current density structure of the cluster.  The permanent spectrum, built up from harder to softer energies over time, should generically be ${\rm d}N_B/{\rm d}\epsilon \propto \epsilon^{-1}$ by the arguments presented in this section.

\subsection{The 256 and 8k runs}

\begin{figure*}
\begin{tabular}{cc}
 \includegraphics[width=80mm]{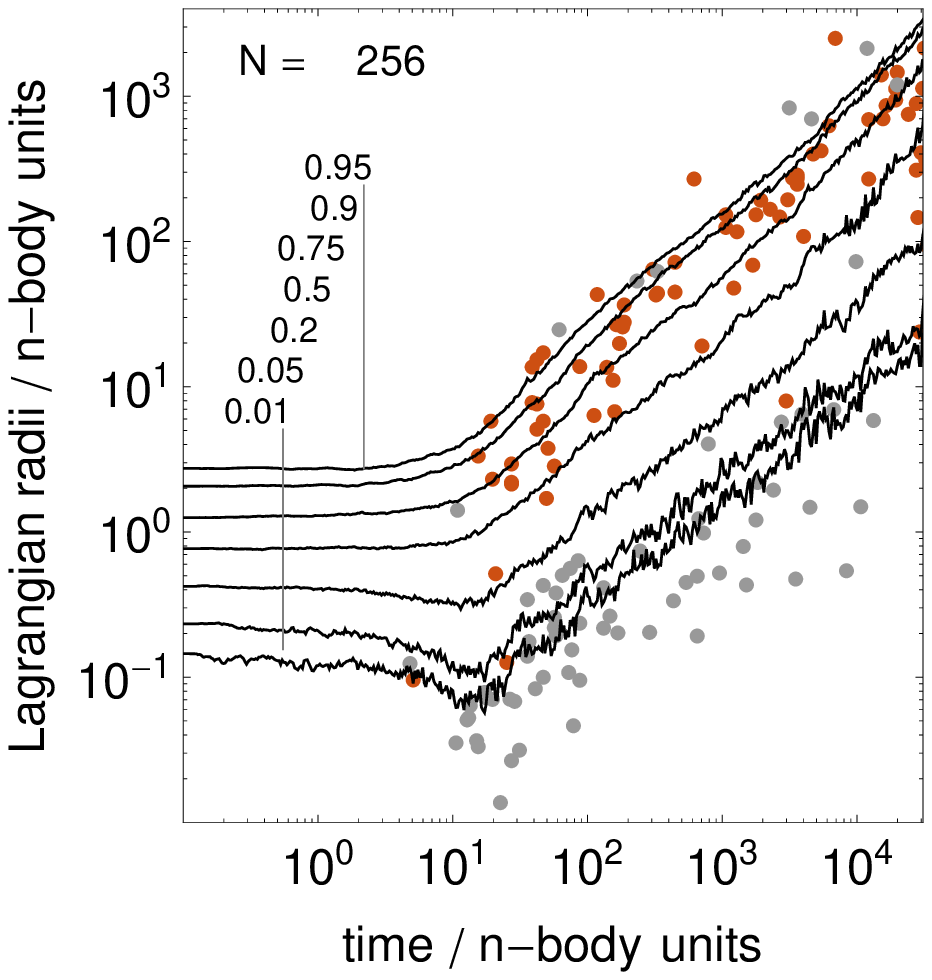}
 \includegraphics[width=80mm]{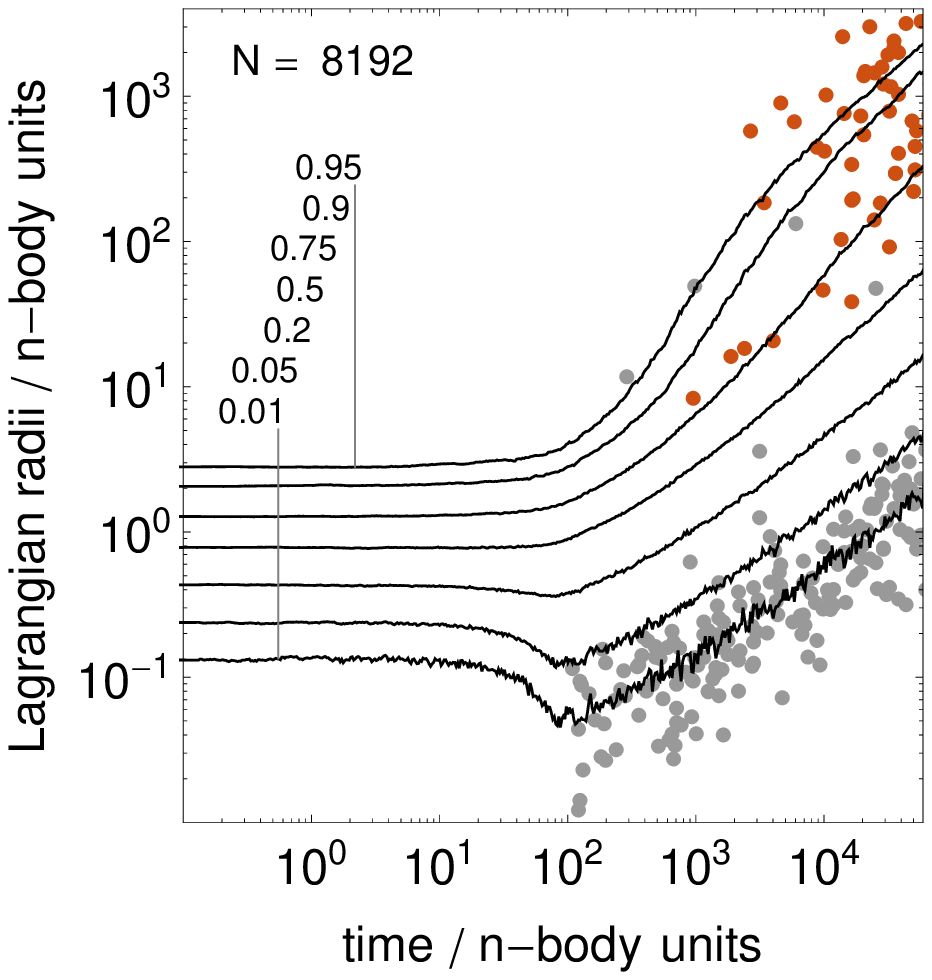}
 \end{tabular}
 \caption{As figure \ref{1024lagr}, but for the 256 and 8k runs.}
 \label{8192lagr}
\end{figure*}

The overall evolution of the 256 and 8k clusters is similar to the 1k case.  In figure \ref{8192lagr} we show the Lagrangian radii and formation sites of binaries, analogous to figure \ref{1024lagr}.  Recall that the 8k runs are analysed at $t=5\times10^4$.  The expansion of the clusters is, as expected, similar to the 1k case modulo a shift of the time of core collapse related to the $N$ dependence of the initial relaxation time, and the formation sites of permanent soft binaries are similarly concentrated exterior to the half-mass radius..  Extending the 8k runs several orders of magnitude in time would have perhaps been ideal, but physical arguments about the relevance of extending the runs (see the discussion below) made the computational expense of that route seem a questionable investment. 

\begin{figure*}
\begin{tabular}{cc}
 \includegraphics[width=80mm]{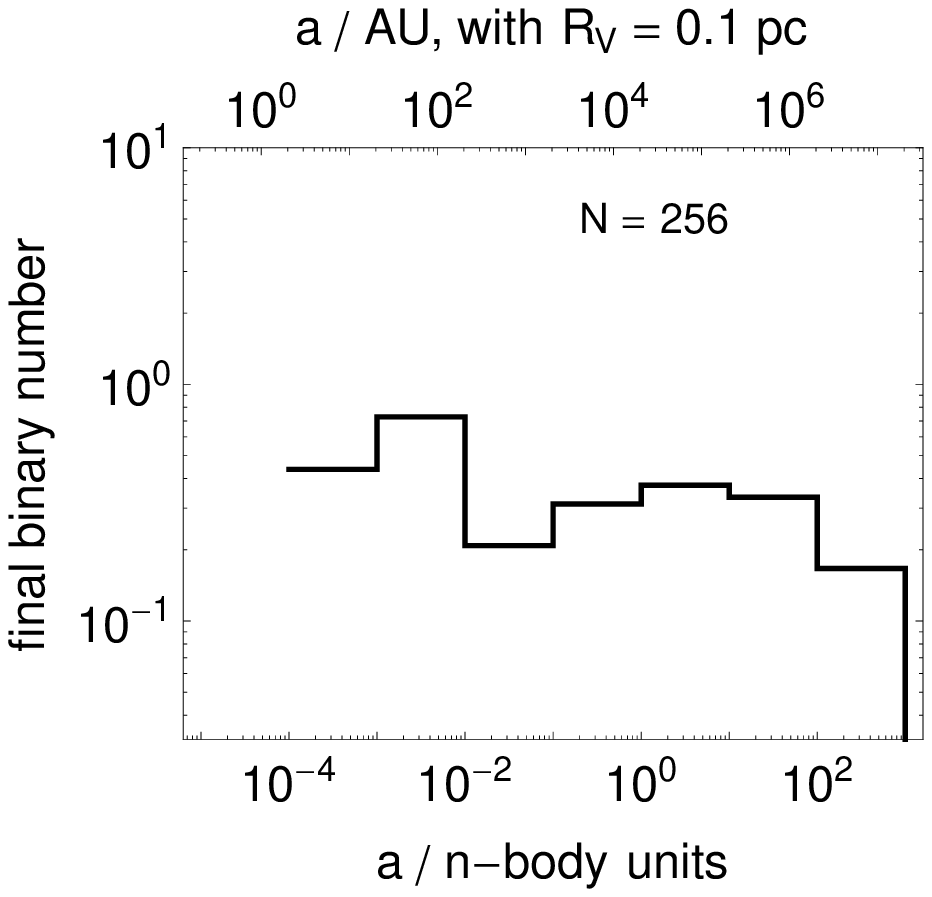} &
 \includegraphics[width=80mm]{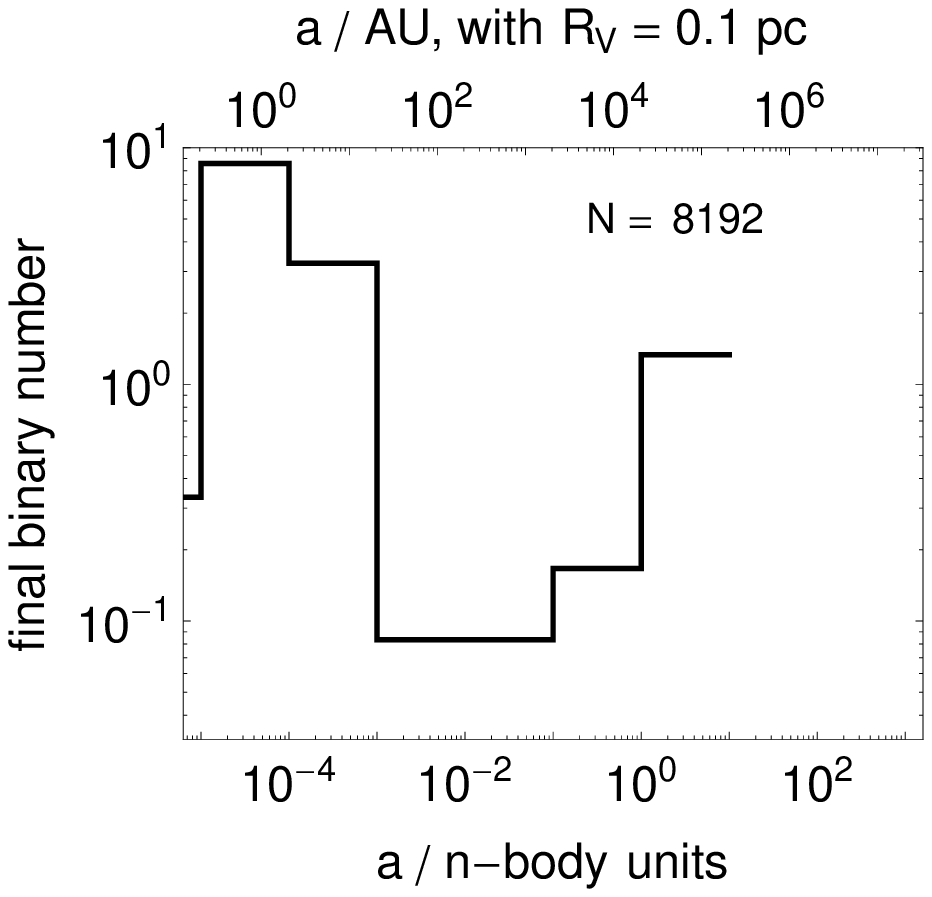} 
\end{tabular}
 \caption{Semi-major axis distribution of the permanent binary populations for the 256 and 8k runs.  These plots are normalized by the total number of runs. The top scale is in physical units with an initial virial radius of 0.1 pc; these can be linearly scaled to any other cluster radius.}
 \label{8192semidist}
\end{figure*}

In figure \ref{8192semidist} we show the semi-major axis distributions for the 256 and 8k runs.  The number of binaries formed at large separations, about one per cluster, are consistent across all the clusters.  The truncation of the distribution for the 8k run is because of the shorter length of (logarithmic) time those clusters were integrated post core-collapse, and because the amount of cluster expansion necessary for the interstellar separation to reach a given value increases with $N^{1/3}$.


\section{Discussion}
\label{conclsection}
We have been dealing with dimensionless units up to this point.  When applying these results to a cluster in a galaxy, there are two limiting length scales to take into account.  First, there is the largest binary that we are interested in producing.  As discussed in the introduction, this is about $10^5$ AU.  Binaries produced above that size scale are not of particular interest because their numbers are depleted by disruption in the galactic field.  Second is the tidal radius of the cluster.  Binaries are produced in the outer regions of the cluster, and as seen in figure \ref{1024lagr} these reach radii of the order 100 \nbody units.  If this is larger than the tidal radius, these binaries will not form, at least not in the relaxation-driven expansion scenario we have discussed.

The first constraint, the upper limit of the binaries we are interested in, means that we are only concerned with binaries with semi-major axes of a less than a few \nbody units, thus cutting off the final two bins of figure \ref{1024semidist}.  This means that each cluster can be expected to produce slightly less than a single binary in the range we are interested in.  The second constraint is that the binary is not formed beyond the tidal radius of the cluster.  With a tidal radius of order 10 pc, our choice of initial virial radius means that the cutoff is of order 100 \nbody units.  Inspecting figure \ref{1024lagr}, we see that binaries continue to be produced closer to the half-mass radius through to the end of the simulation inside of 100 \nbody units, but the production in the outskirts will only continue in a truly and unrealistically isolated cluster.  

These two constraints both cut off approximately the same binaries from consideration, however. This is to be expected, to order of magnitude.  Roughly speaking, the cluster becomes tidally limited as its mean density reaches some critical value, and a binary in the field can be thought of as limited by a similar density condition. Since the binaries are formed at about the interstellar separation, it follows that the two constraints affect the same binaries.  As is clear from figure \ref{1024freeze}, after the outer Lagrangian radii are outside a realistic tidal limit at a few thousand time units, the bulk of the binaries with separations less than 10 \nbody units have already been created, and the story is the same for the 256 runs.

Looking to the 8k runs, the situation is somewhat grimmer on the soft-binary production front.  The formation site of nearly all the binaries are at radii beyond several hundred \nbody units, flirting with the tidal radius with the same initial scaling of $R_v=0.1$ pc.  A perhaps more-reasonable initial virial radius for this larger cluster of, say, 0.25 pc exacerbates the problem; more populous clusters need to be in a tidally unrestricted setting in order to expand sufficiently to produce the soft binaries we are interested in.  It appears that when the constraints of the galaxy are included, it is the less-populous clusters that are more likely to contribute to the population of wide binaries, for two reasons.  First, with each cluster producing of order a single wide binary, 10 dispersed clusters of 200 stars will yield more wide field binaries than a single ONC of about 2000.  Second, the tidal limitation restricts the ability of larger clusters to produce even their single wide binary.

In the most recent comprehensive survey of multiplicity in the Solar neighbourhood, encompassing 454 solar-type stars within 25 pc, \citet{raghavan10} find 10 binaries with estimated semimajor axes greater than $10^4$ AU. Thus locally, $\sim 2$\% of field stars are in the very wide systems that we have explored here. With approximately one such binary being created from each dissolving cluster, a field populated by 1k clusters would have a 0.1\% wide binary population, while in a field populated by our 256 star clusters it would be more like 1\%. The statistics of the local field are thus broadly consistent with the field being synthesized by clusters of a few hundred stars, rather than a few thousand. This matches the statistics of clustered star formation sites \citep{lada03,porras03}, which appear to have a median cluster membership of $\sim 300$ \citep{adams06}. We emphasize that the clustered formation scenario should be thought of in terms of a hierarchical structure rather than as clustered versus isolated modes \citep{bressert10}.

This formation process relies upon relaxation-driven expansion.  Before this point of a cluster's life, there should not be any wide binaries.  This is consistent with the lack of observed wide binaries in the ONC \citep{scally99}, and it is possible that a small number of wide binaries might be formed subsequently during the dissolution of the ONC. It has often been noted that the observed binary statistics in the ONC are similar to that of the field (thus encouraging the notion that the field might be comprised of dissolved ONC-type clusters), with the notable exception of the fact that the ONC is lacking in wide binaries. Conceivably, therefore, this problem could be solved if the wide binaries are yet to form. It is however worth emphasizing that the current dynamical status of the ONC is not well known observationally and that, strictly speaking, our present analysis should not be quantitatively applied to clusters that expand from super-virial conditions as a result of gas expulsion.

We have not included any primordial binaries in these simulations. As \citet{kouwenhoven10} point out, the effect of primordial binarity depends on the characteristics of those binaries. The proper way to set up a primordial population is not obvious, especially when considering the very early evolution of a cluster where some studies advocate significant early dynamical processing of the primordial population \citep[e.g.][]{kroupa95,parker09a}, and others argue that the binary population is largely stable by the time gravitational dynamics become the dominant physical process \citep{moeckel10}. A full treatment of this issue is beyond the scope of this paper, but we can draw on previous work to speculate on the likely effects.

\citet{kouwenhoven10} find that including a primordial binary population can lead to hierarchical systems consisting of a primordial binary and a dynamicall-formed very wide companion. If the primordial binary is small enough that it can approximately be treated as its center of mass, i.e. it is energetically hard by local standards, then we expect that some fraction of the wide systems produced would be members of hierarchical triples (or possibly quadruples). Interactions between more comparably sized binaries will naturally be more complex; these will be sensitive to the length of time the cluster spends pre-core collapse, since this determines the amount of processing the primordial binaries may undergo. This timescale is highly dependent on the initial density structure and mass function of the cluster, and this is perhaps an avenue for further work.

Regarding the initial cluster structure, we note that while we have used a Plummer model as our initial density distribution, any standard cluster setup should yield very similar post-collapse structures. \citet{baumgardt02} show by way of example that after core collapse, a Plummer sphere and a $W_0 = 3$ \citet{king66} model develop density profiles that are very well matched, provided the time of comparison is large compared to the collapse timescale. We expect that the outer extent of any expanding cluster should satisfy the conditions to produce very wide binaries. Due to the nearly self-similar post-collapse evolution, clusters with extremely high initial densities will evolve through the binary-creating stages we simulated, provided the tidal limit allows the interstellar separation to reach the requisite large value to accommodate the binaries. We further expect that subvirial, fractal, or other alternate cluster setups should proceed similarly, as the expansion after the densest phase of core collapse is driven by the same physics.

Finally, we note that the number of wide binaries we find in these simulations (of order 0.1\% for the 1k runs) is smaller by roughly an order of magnitude than found by \citet{kouwenhoven10} and \citet{moeckel10} (about 1\%).  This is due to our insistance on the long-term permanence of the binaries, which we check by integrating the simulations a factor of two longer than the point at which we analyse them.  Including all of the instantaneous binaries in our analysis boosts the numbers to consistency with those previous studies.

\section{Conclusions}
\label{conclusions}
In an effort to identify the physics behind the formation of very wide binaries (semi major axes $\sim 10^4 - 10^5$ AU), we have performed simulations of isolated star clusters undergoing relaxation-driven expansion after core collapse. Our explanation for the formation of these very soft, yet permanent binaries centers around the effect of the changing cluster structure on the statistical mechanics that are constantly creating and destroying binaries at all energies. Specifically, when the local density drops on a timescale short compared to the disruption time of a binary, it can be frozen out of the statistical equilibrium and become permanent.

The resultant permanent binary population has some interesting characteristics. Namely, the binaries form in the outskirts of the cluster, with semi-major axes at approximately the local interstellar separation; the final semi-major axis distribution is roughly flat in log($a$); and the total number of binaries formed by any single cluster is of order unity, regardless of the population of the cluster. Since approximately 2\% of local field stars are in binaries with separations greater than $10^4$ AU, the production mechanism presented here could account for that fraction if the field is composed of dissolved clusters with a characteristic population of a few hundred stars, which is consistent with the statistics of clustered star formation sites.

\section*{Acknowledgments}
Our thanks to Douglas Heggie for his comments on the manuscript, to the referee for a helpful report, and to Sverre Aarseth for generously sharing his expertise and his GPUs.


\appendix
\section{Calculating the Binary Freeze-out Separation}
\label{FreezeOutAppendix}

Because the cluster is expanding, it is not in the homogeneous state from which the steady-state distribution is derived by \citet{goodman93a}.  If we place ourselves in the shoes of a given soft binary, we have some expected lifetime that is a function of our energy, the local stellar density, and the local velocity dispersion \citep{heggie75}.  In an expanding cluster, this expected lifetime is a constantly changing quantity, and may in some cases become longer than the cluster expansion timescale on which the local density declines. As the cluster expands, progressively larger-separation binaries are thus frozen-out from the binary creation-destruction cycle.  Binaries that are energetically soft by their local standards are then able to survive for long times.

\begin{figure}
 \includegraphics[width=80mm]{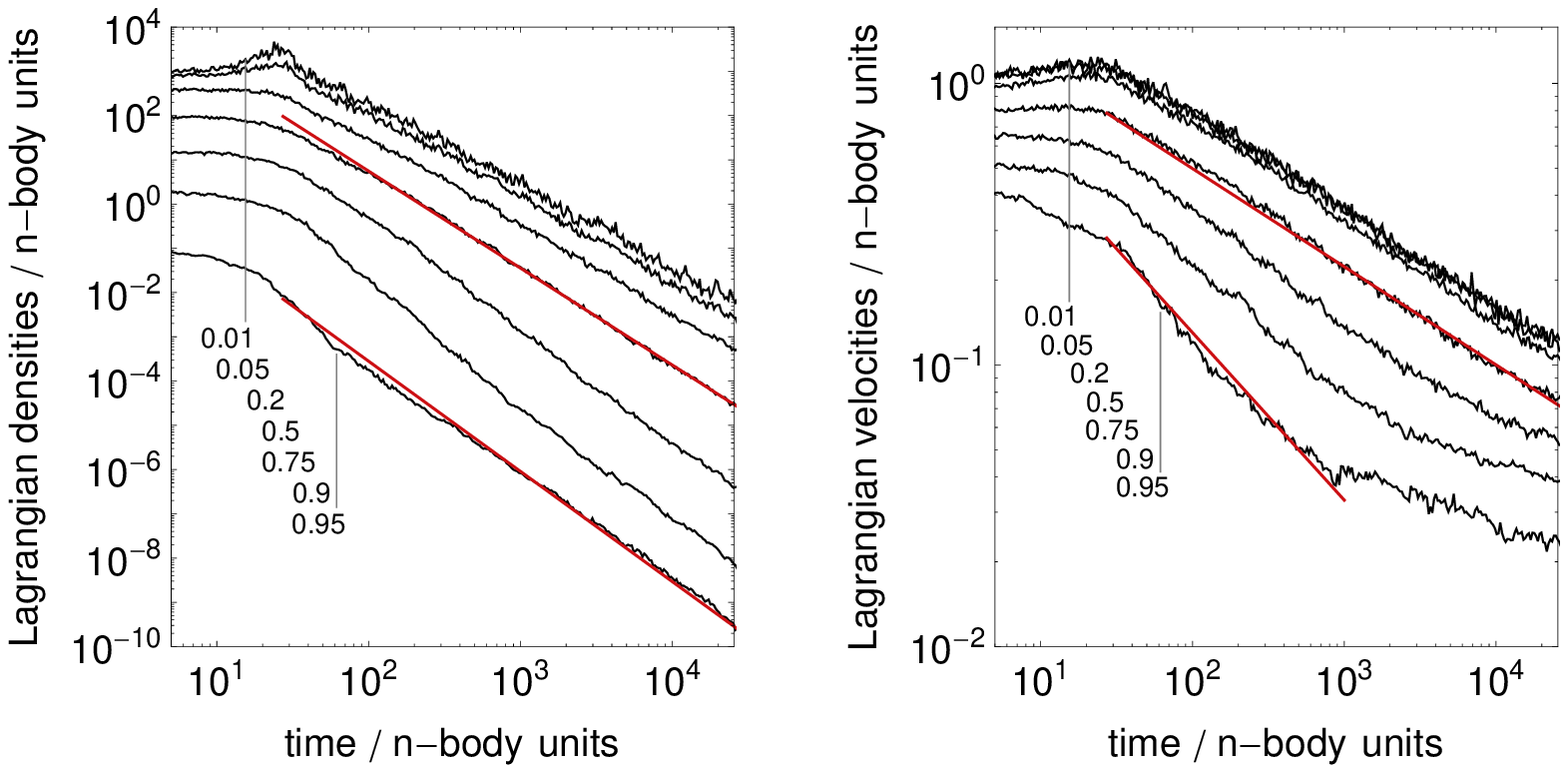}
 \caption{Lagrangian densities and velocity dispersions for the 1k clusters, calculated as described in the text.  The red lines are equations \ref{densevolution} and \ref{dispevolution}.}
 \label{1024densdisp}
\end{figure}

Given the power-law expansion of the Lagrangian radii given by equation \ref{rlevolution}, we can make predictions of this freeze-out process.  In what follows, for simplicity we take all stars to be the mean mass in the cluster.  The density at a Lagrangian radius (we will call this a `Lagrangian density') that follows equation \ref{rlevolution} will evolve as 
\begin{equation}
n_{L} = n_{L,0} \left(1 + \chi \frac{t -t _{cc}}{t_{rh,0}} \right)^{-\alpha},
\label{densevolution}
\end{equation}
with the exponents related by $\alpha = 3\delta$.  We empirically find a similar relation for the velocity dispersion at a Lagrangian radius, so that 
\begin{equation}
\sigma_{L} = \sigma_{L,0} \left(1 + \chi \frac{t -t _{cc}}{t_{rh,0}} \right)^{-\beta}.
\label{dispevolution}
\end{equation}
When calculating the Lagrangian dispersion at a given radius, we consider stars inside a shell of width 0.02 around that radius.  For example, the dispersion at the 0.95 radius is calculated using the stars between the 0.94 and 0.96 radii.  We subtracted off the mean radial velocity of the shell, so that the bulk expansion of the cluster does not figure in to any calculations of local quantities.  The evolution of the Lagrangian densities and dispersions are shown in figure \ref{1024densdisp}, as well as the fits given by equations \ref{densevolution} and \ref{dispevolution} for the 0.5 and 0.95 Lagrangian radii.  In practice we normalized the fits to produce the best match, and we only use them where it is visually reasonable (thus the limited range of the fit to the 0.95 dispersion in the 1k run).

The destruction rate for a soft binary at a given energy is calculated by \citet{heggie75}, and can be cast into a rate for a given semi-major axis and approximated as 
\begin{equation}
Q_a^- = \frac{20\sqrt{2 \pi}}{3\sigma} G m a.
\end{equation}

Collecting constant terms as $\lambda$, we can write the evolution of a number of binaries at that semi-major axis as 
\begin{align}
\dot{N_a} & = - n Q_a^- N_a\\
&= - \lambda N_a \left( 1 + \chi \frac{t - t_{cc}}{t_{rh,0}} \right)^{-\alpha + \beta}.
\end{align}

The solution to this equation for times greater than some time $t^\star$ is
\begin{align}
{\rm ln} & N_a(t) = N_a(t^\star)
\frac{\lambda t_{rh,0}}{\chi \left( 1-\alpha+\beta \right) } \times \\
\times&\left[
\left(1+\chi \frac{t^\star-t_{cc}}{t_{rh,0}}\right)^{1-\alpha+\beta}
-\left(1+\chi \frac{t-t_{cc}}{t_{rh,0}}\right)^{1-\alpha+\beta} 
\right].
\end{align}
For $1-\alpha+\beta<0$ the solution is, roughly speaking, an asymptotic limit at $t \rightarrow \infty$ (the first term in square brackets) and a decay to that limit from the initial population (the second term).  The timescale for decay is something like $t_{rh,0}$, which is short compared to length of the simulation, so the behavior of the asymptotic limit appears to be the most interesting feature.  

We define the time when a population of binaries at some semi-major axis is frozen out as the time when their asymptotic limit is $e^{-1}$.  While this is somewhat arbitrary, other reasonable choices (such as one-half surviving) introduce only a small correction in what follows.  Reintroducing the numerical factors contained in $\lambda$, we find that for a Lagrangian shell with initial density $n_0$ and velocity dispersion $\sigma_0$ the largest semi-major axis that is frozen out evolves as 

\begin{equation}
a_{frz}(t) = \left( \frac{-3 \chi \sigma_0(1-\alpha+\beta)}{20 \sqrt{2 \pi} n_0 G m t_{rh,0}} \right) \left(1 + \chi \frac{t-t_{cc}}{t_rh,0} \right)^{1-\alpha+\beta}.
\label{freezevalue}
\end{equation}
This calculation is not a guarantee that a given binary will survive, but it provides an estimate of when we might expect to see binaries at a given semi-major axis start to become permanent at different Lagrangian radii.

\section{The Statistical Binary Energy Distribution in a Non-Uniform Density Field}
\label{DensityAppendix}
At very soft energies, the equilibrium distribution of binary energies found by \citet{goodman93a} converges to a power law with ${\rm d}N_B/{\rm d}\epsilon \propto \epsilon^{-5/2}$, a consequence of a Maxwellian distribution of velocities and a uniform number density \citep{heggie75}. Here we illustrate the effect of a non-uniform density distribution, where we assume that the typical star sees a number density that is a power law with radius, $n(r) \propto r^{-q}$.

The absolute value of the specific binding energy of two stars of mass $m$, separation $r$ and relative velocity $v$ is 
\begin{equation*}
\epsilon = \frac{2 G m}{r} - \frac{1}{2}v^2 \equiv x - \frac{1}{2}v^2,
\end{equation*}
where we have introduced $x$ for the gravitational energy of the pair. The number density can then be written as $n(r) \propto x^{q}$. With a Maxwellian distribution of relative velocities with dispersion $\sigma$, the number of potential binary partners a star sees in the range $\epsilon$ to $\epsilon + {\rm d} \epsilon$ is 
\begin{equation*}
N_\epsilon \propto \left(\int_{Gm/\epsilon}^{\infty} \left . \frac{\partial v}{\partial \epsilon} \right | _r  v^2 {\rm exp} \left ( -\frac{v^2}{2 \sigma^2} \right) r^2 n(r) {\rm d} r \right) {\rm d}\epsilon,
\end{equation*}
with the lower bound on the integral over all radii given by the zero-velocity binary pair at energy $\epsilon$. Note that $n(r)$ is not the density structure of the cluster, but rather the density profile seen by a star, weighted by all the stars in the cluster.  Our assumption is that this averaged profile is fit by a power law over some range. For a Plummer sphere, for instance, $q \sim 1$ is appropriate for $r$ comparable to the length scale of the distribution.

Converting from radius to $x$ and including the power-law density assumption, we have
\begin{equation*}
N_\epsilon \propto \left (  \int_{\epsilon}^{\infty} \left( x - \epsilon \right)^{1/2} {\rm exp} \left ( \frac{\epsilon - x}{2 \sigma^2} \right) x^{q-4} {\rm d} x \right ) {\rm d}\epsilon
\end{equation*}

We now introduce $y \equiv x - \epsilon$, and rewrite the integral as 
\begin{equation*}
N_\epsilon \propto  \left( \int_{0}^{\infty} y^{1/2} {\rm exp} \left ( \frac{-y}{2 \sigma^2} \right) (y+\epsilon)^{q-4} {\rm d} y \right) {\rm d}\epsilon.
\end{equation*}

The exponential term limits the contribution to the integral to values of $y \lesssim \sigma^2$,  and since we are concerned with very soft energies we have $\epsilon \ll \sigma^2$.  We can consider the behavior of the integrand in two limits.  When $\sigma^2 \gg y \gg \epsilon$, the integrand is approximately $y^{q-7/2}$.  When $\sigma^2 \gg \epsilon\gg y$, the limiting behavior of the integrand is $y^{1/2}$.  Thus for $q \ll 7/2$, the dominant contribution to the integral occurs with values $y \sim \epsilon$.   At this point the integrand is turning over to the $y \gg \epsilon$ limit,  $y^{q-7/2}$, and the integral is approximately 
\begin{equation*}
N_\epsilon \propto \epsilon^{q-5/2} {\rm d}\epsilon.
\end{equation*}  
While this is approximate due to the unpalatable nature of the integral, it recovers the uniform density result when $q=0$.  As $q$ increases the peak of the integrand broadens, and this treatment becomes troublesome.  However, empirically a Plummer sphere yields $q \sim 1$, and we feel comfortable estimating the integral in this fashion.  With $q \sim 1$, we find the approximate ${\rm d}N_B/{\rm d}\epsilon \propto \epsilon^{-1.5}$ seen in our statistical spectrum over a significant range in figure \ref{1024semidist}.

\bsp

\label{lastpage}

\end{document}